\documentclass{acm_proc_article-sp}

\usepackage{graphicx}
\graphicspath{ {} }
\usepackage{float}

\usepackage{color}

\usepackage{etoolbox}
\makeatletter
\patchcmd{\maketitle}{\@copyrightspace}{}{}{}
\makeatother

\begin{document}

\title{Privacy Implications of Health Information Seeking on the Web}
\subtitle{\emph{Communications of the ACM, March 2015}}
\numberofauthors{1}

\author{
\alignauthor
Timothy Libert\\
       \affaddr{University of Pennsylvania}\\
       \affaddr{3620 Walnut Street}\\
       \affaddr{Philadelphia, PA, USA}\\
       \email{tlibert@asc.upenn.edu}
}

\maketitle
\begin{abstract}
This article investigates privacy risks to those visiting health-related web pages.  The population of pages analyzed is derived from the 50 top search results for 1,986 common diseases.  This yielded a total population of 80,124 unique pages which were analyzed for the presence of third-party HTTP requests.  91\% of pages were found to make requests to third parties.  Investigation of URIs revealed that 70\% of HTTP Referer strings contained information exposing specific conditions, treatments, and diseases.  This presents a risk to users in the form of personal identification and blind discrimination.  An examination of extant government and corporate policies reveals that users are insufficiently protected from such risks.
\end{abstract}

% A category with the (minimum) three required fields
%\category{H.4}{Information Systems Applications}{Miscellaneous}
%A category including the fourth, optional field follows...
%\category{D.2.8}{Software Engineering}{Metrics}[complexity measures, performance measures]

\terms{Privacy, Health, Web, Tracking}

%\keywords{ACM proceedings, \LaTeX, text tagging} % NOT required for Proceedings

\section{Introduction}

	Privacy online is an increasingly popular field of study, yet it remains poorly defined.  ``Privacy" itself is a word which changes according to location, context, and culture.  Additionally, the web is a vast landscape of specialized sites and activities which may only apply to a minority of users - making defining widely-shared privacy concerns difficult.  Likewise, as technologies and services proliferate, the line between on- and off-line is increasingly blurred.  Researchers attempting to make sense of this rapidly changing environment are frequently stymied by such factors.  Therefore, the ideal object of study is one which is inherently sensitive in nature, applies to the majority of users, and readily lends itself to analysis.  The study of health privacy on the web meets all of these criteria.

	Health information has been regarded as sensitive since the time of the ancient Greeks.  In the 5th century B.C., physicians taking the Hippocratic Oath were required to swear that: \emph{Whatever I see or hear in the lives of my patients...I will keep secret, as considering all such things to be private}.\cite{nih-2002-hippocratic}  This oath is still in use today, and the importance of health privacy remains universally recognized. However, as health information seeking has moved online, the privacy of a doctor's office has been traded in for the silent intrusion of behavioral tracking.  This tracking provides a valuable vantage point from which to observe how established cultural norms and technological innovations are at odds.

	Online health privacy is an issue which affects the majority of Internet users.  According to the Pew Research Center, 72\% of adult Internet users in the U.S. go online to learn about medical conditions.\cite{pew-2013-health}  Yet only 13\% of these begin their search at health-specific sites.  In fact, health information may be found on a wide spectrum of sites ranging from newspapers, discussion forums, to research institutions.  In order to discover the full range of sites users may visit when seeking health information, I used a search engine to identify 80,142 unique health-related web pages by compiling responses to queries for 1,986 common diseases.  This selection of pages represents what users are \emph{actually} visiting, rather than a handful of specific health portals.

	Having identified a population of health-related web pages, I created a custom software platform to monitor the HTTP requests initiated to third-parties.  I discovered that 91\% of pages make requests to additional parties, potentially putting user privacy at risk.  Given that HTTP requests often include the URI of the page currently being viewed (known as the ``Referer" [sic]), information about specific symptoms, treatments, and diseases may be transmitted.  My analysis shows that 70\% of URIs contain such sensitive information.

	This proliferation of third-party requests makes it possible for corporations to assemble dossiers on the health conditions of unwitting users.  In order to identify which corporations are the recipients of this data I have also analyzed the ownership of the most requested third-party domains.  This has produced a revealing picture of how personal health information becomes the property of private corporations.

	This article begins with a short primer on how third-party HTTP requests work, moves on to previous research in this area, details methodology and findings, and concludes with suggestions for protecting health privacy online.

\section{Background: Third-Party HTTP Requests}
	
	% FIGURES
	\begin{figure*}
		\centering
		\includegraphics[scale=.23]{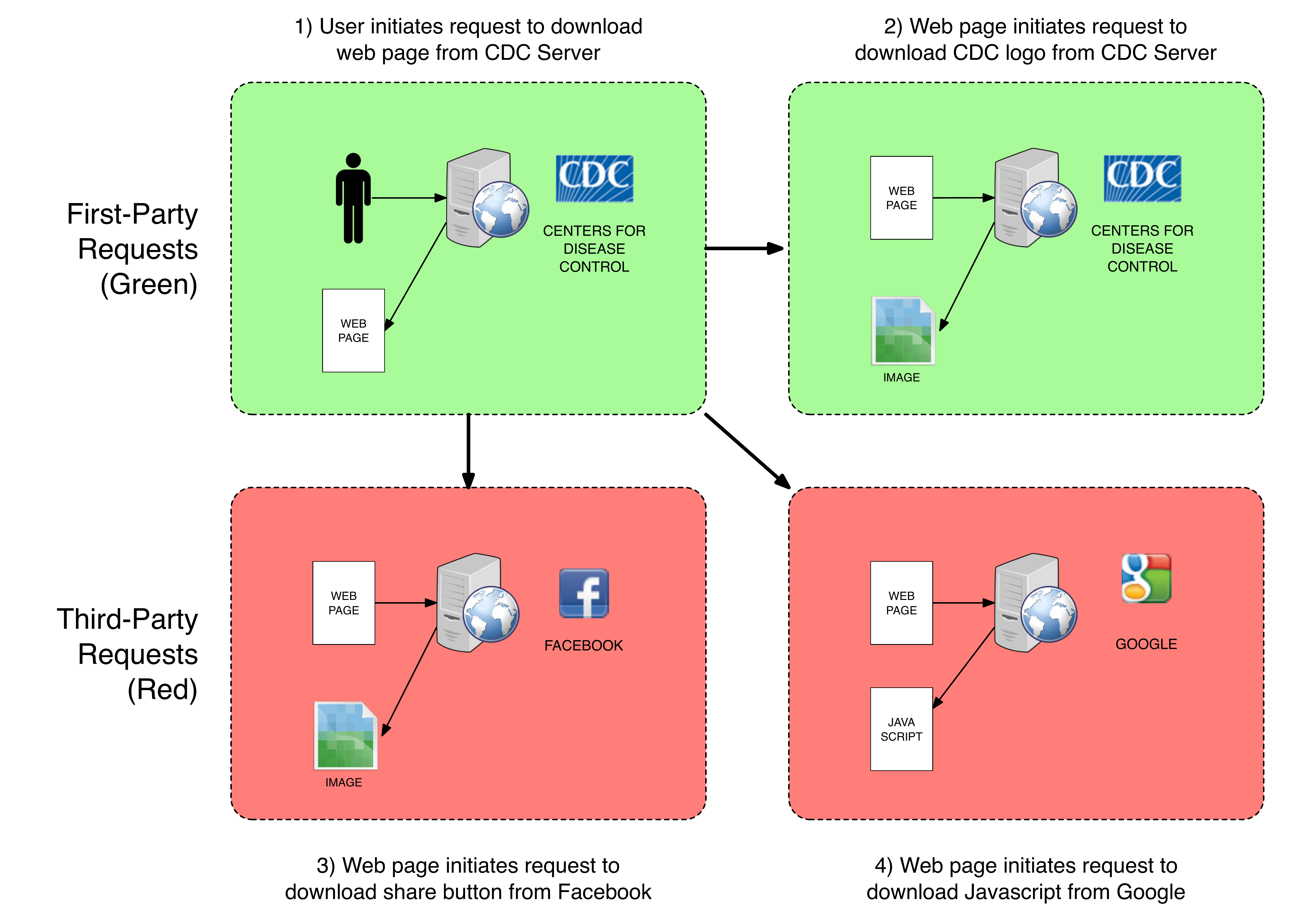}
		\caption{First- and Third-Party Requests on the CDC Web Page for HIV/AIDS}
	\end{figure*}
	
	A real-world example is the best way to understand how the information is leaked to third-parties on a typical web page.  When a user searches online for ``HIV" one of the top results is for the U.S. Centers for Disease Control and Prevention (CDC) page with the address ``http://www.cdc.gov/hiv/''\footnote{As of April, 2014}.  Clicking on this result initiates what is known as a ``first-party'' Hypertext Transfer Protocol (HTTP) request to the CDC webserver (Figure 1.1).  A portion of such a request is as follows:

	\begin{verbatim}
		GET /hiv/
		Host: www.cdc.gov
		User-Agent: Mozilla/5.0 (Macintosh...
	\end{verbatim}
	
	This request is sent to the CDC webserver (``Host: www.cdc.gov") and is an instruction to return (``GET") the page with the address ``/hiv/".  This request also includes ``User-Agent" information which tells the server what kind of browser and computer the user has.  In this case, the user employs the Mozilla Firefox browser on a Macintosh computer.  Such information is helpful when loading specially-optimized pages for smartphones or tablets.  
	
	Once this request has been made, the CDC webserver sends the user an HTML file.  This file contains the text of the page as well as a set of instructions that tells the web browser how to download and style additional elements such as images (Figure 1.2).  In order to get the CDC logo, the following HTTP request is made:
	
	\begin{verbatim}
		GET /TemplatePackage/images/cdcHeaderLogo.gif
		Host: www.cdc.gov
		User-Agent: Mozilla/5.0 (Macintosh...
		Referer: http://www.cdc.gov/hiv/
	\end{verbatim}
		
		This request introduces a new piece of information called the ``Referer".  The ``Referer" contains the address of the page the user is currently viewing.  The CDC web server may keep records of all HTTP requests in order to determine what pages and content are being requested most often.
		
		Because the ``Host" for both requests is identical (www.cdc.gov), the user is only interacting with a single party and such requests are called ``first-party requests".  The only two parties who know that the user is looking up information about HIV are the user and the CDC.  However, the HTML file also contains code which makes requests to outside parties.  These types of ``third-party requests" typically download ``third-party elements" such as images and javascript.  Due to the fact that users are often unaware of such requests, they form of the basis of the so-called ``invisible web".
		
		On the CDC's HIV page, third-party requests are made to the servers of Facebook, Pinterest, Twitter, and Google.  In the case of the first three companies, the requested elements are all social media buttons which allow for the sharing of content via the ``Recommend", ``Tweet", or ``Pin It" icons (Figure 1.3).  It is unlikely that many users would understand that the presence of these buttons indicates that their data is sent to these companies.  In contrast, the Google elements on the page are entirely invisible and there is no Google logo present.  One of these requests is sent to Google's Analytics service (Figure 1.4) to download a file containing Javascript code:
	
	\begin{verbatim}
		GET /ga.js
		Host: www.google-analytics.com
		User-Agent: Mozilla/5.0 (Macintosh...
		Referer: http://www.cdc.gov/hiv/
	\end{verbatim}
	
	Again, the ``Referer" field reveals that the user is visiting a page about HIV.  By pairing information about the User-Agent, Referer, and user's IP address, it is possible for companies like Google and Facebook to identify people who are concerned with HIV.\cite{yen-2012-host}  In all likelihood those visiting this page are unaware of this fact, and would not be happy to find out.

\section{Prior Research}
	 Prior research has demonstrated that while users are uncomfortable with this type of tracking, it is performed in a number of highly sophisticated ways, and it is increasingly 	widespread.

	\subsection{Attitudes}
	There has long been anxiety about how personal data will be used on the web.  A 1999 study determined that ``only 13\% of respondents reported they were `not very' or `not at all' concerned" about their privacy online.\cite{ackerman-1999-privacy}  Such anxiety remained in 2003 when 70\% of survey respondents reported that they were nervous that websites had information about them.\cite{turow-2003-americans}  A 2009 follow-up study revealed that 67\% of respondents agreed with the statement that they had ``lost all control over how personal information is collected and used by companies".\cite{turow-2009-americans}  These surveys demonstrate that the activities of many businesses run directly counter to public preferences.

	As with general concerns with online privacy, there is excellent research exploring attitudes towards health information.  In 2012, Hoofnagle et al determined that only 36\% of survey respondents knew that advertisers are allowed to track their visits to health-related websites.\cite{hoofnagle-2012-privacy}  An extensive study from the year 2000 found that 85\% of Internet users in poor health were concerned that websites would share their data, and only 3\% were comfortable with websites sharing their data with ``other sites, companies, and advertisers".\cite{grimes-2000-ethics}  Despite these fears, 44\% of respondents felt that their information was safe with institutions such as the National Institutes of Health (NIH).\cite{grimes-2000-ethics}  The above CDC example indicates that this trust is potentially misplaced.

	\subsection{Mechanisms}
	Once a third-party request is made, a user may be tracked using a number of ever-evolving technical mechanisms.  Researchers have been tracing the development of such mechanisms for years, often analyzing the code and behaviors which take place within the web browser.  These are often called ``client-side" techniques for they take place on the user's computer.  Traditional client-side techniques typically involve storing data on the user's computer in small text files known as ``cookies" - this functions as a sort of digital name-tag.\cite{ayenson-2011-flash}  Users are getting more adept at evading such practices, therefore newer techniques often employ so-called ``browser fingerprinting" to identify users based on characteristics of their computers.  This area of research has proven very popular as of late with numerous studies investigating fingerprinting techniques.\cite{jackson-2006-protecting}\cite{eckersley-2010-unique}\cite{jang-2010-empirical}\cite{nikiforakis-2013-cookieless}\cite{acar-2013-fpdetective}  In addition, Miller et al have recently demonstrated sophisticated attacks on HTTPS which are able to reveal ``personal details including medical conditions."\cite{miller-2014-know}

	Turning attention to the server-side, Yen et al have recently demonstrated a tracking technique which utilizes a combination of IP address, User-Agent string, and time intervals when HTTP requests were made.  This team was able identify users 80\% of the time, which is on par with what is typically accomplished with client-side cookies.\cite{yen-2012-host}  Furthermore, identification rates remained essentially static even when removing the final octet of the IP address, which is a common technique by which major advertisers claim to anonymize data.  Yen et al's findings indicate that while novel techniques may be needed on the client-side, the lowly HTTP request is sufficient for advanced server-side techniques.

	\subsection{Measurement}
	The final area of related research is measurement.  Measurement of web tracking generally entails two steps: selecting a population of pages, and performing automated analysis of how user data is transmitted to third parties.  Many studies have relied on ``popular site" lists provided by the Alexa company\cite{krishnamurthy-2006-generating}\cite{Castellucia-2013-dataharvesting2}\cite{krishnamurthy-2009-privacy}\cite{roesner-2012-detecting}, but often utilize their own methodologies for analysis.  Krishnamurthy and Wills have conducted many of the most important studies in this area\cite{krishnamurthy-2006-generating} and developed the idea of a ``privacy footprint"\cite{krishnamurthy-2009-privacy} which is based upon the number of ``nodes" a given user is exposed to as they surf the web.  This team has consistently found that there are high levels of tracking on the web, including on sites dealing with sensitive personal information such as health.\cite{krishnamurthy-2009-privacy}  Other teams have performed comparative analyses between countries\cite{Castellucia-2013-dataharvesting2} as well as explored general trends in tracking mechanisms.\cite{mayer-2012-third}\cite{roesner-2012-detecting}  A common theme among all measurement research is that the amount of tracking on the web is increasing, and shows no signs of abating.  The data presented in this article updates and advances extant findings with a focus on how users are tracked when they seek health information online. 

\section{Methodology}
	In order to quickly and accurately reveal third-party HTTP requests on health-related web pages, my methodology has four main components: page selection, third-party request detection, request analysis, and corporate ownership analysis.

	\subsection{Page Selection}
	
	A variety of websites such as newspapers, government agencies, and academic institutions provide health information online.  Thus, limiting analysis to popular health-centric sites fails to reach many of the sites users actually visit.\cite{krishnamurthy-2011-privacyleakage}  To wit, the Pew Internet and American Life Project found that ``77\% of online health seekers say they began at a search engine such as Google, Bing, or Yahoo"\cite{pew-2013-health} as opposed to a health portal like WebMD.com.  In order to best model the pages a user would visit after receiving a medical diagnosis, I first compiled a list of 1,986 diseases and conditions based on data from the Centers for Disease Control, the Mayo Clinic, and Wikipedia.  Next, I used the Bing search API in order to find the top 50 search results for each term.\footnote{Search results were localized to US/English.}  Once duplicates and binary files (pdf, doc, xls) were filtered out, a set of 80,142 unique web pages remained.  A major contribution of this study to prior work is the fact that my analysis is focused on the pages which users seeking medical information are most likely to visit, irrespective of if the site is health-centric.

	\subsection{Third-Party Request Detection}

	To detect third-party HTTP requests, my methodology employs a ``headless'' web browser named PhantomJS.\cite{phantomjs-2013-website}  PhantomJS requires no GUI, has very low resource utilization, and is therefore well suited for large-scale analyses.  Due to the fact that it is built on WebKit, PhantomJS's underlying rendering engine is capable of executing Javascript, setting and storing cookies, and producing screen captures.  Most important for this project, PhantomJS allows for the direct monitoring of HTTP requests without the need to resort to browser hacks or network proxies.
	
	It should be noted that the most recent versions of PhantomJS (1.5+) do not support the Adobe Flash browser plug-in.  To address this potential limitation, I conducted testing with an older version of PhantomJS (1.4) and Flash.  The inclusion of Flash led to much higher resource utilization, instability, and introduced a large performance penalty.  While this method successfully analyzed Flash requests, I determined that Flash elements were comparatively rare and had negligible effect on the top-level trends presented below.  Therefore, I made the decision to forgo analysis of Flash requests in favor of greater software reliability by using the most recent version of PhantomJS (1.9).

	In order to fully leverage the power of PhantomJS, I created a custom software platform named \emph{webXray} which drives PhantonJS, collects and analyzes the output in Python, and stores results in MySQL.  The workflow begins with a pre-defined list of web page addresses which are ingested by a Python script.  PhantomJS then loads the given web address, waits 30 seconds to allow for all redirects and content loading to complete, and sends back JSON-formatted output to Python for analysis.  This technique represents an improvement over methods such as searching for known advertising elements detected by popular programs such as Ghostery or AdBlock.\cite{Castellucia-2013-dataharvesting2}  As of March 2014, Ghostery reports that the WebMD web page for ``HIV/AIDS" contains four trackers.  In contrast, \emph{webXray} detects the same page initiating requests to thirteen distinct third-party domains.  This is due to the fact that Ghostery and AdBlock rely on curated ``blacklists" of known trackers, rather than reporting \emph{all} requests.
	
	% really want this by the next section, but have to put it here b/c latex!
	\begin{figure*}
		\centering
		\includegraphics[scale=.5]{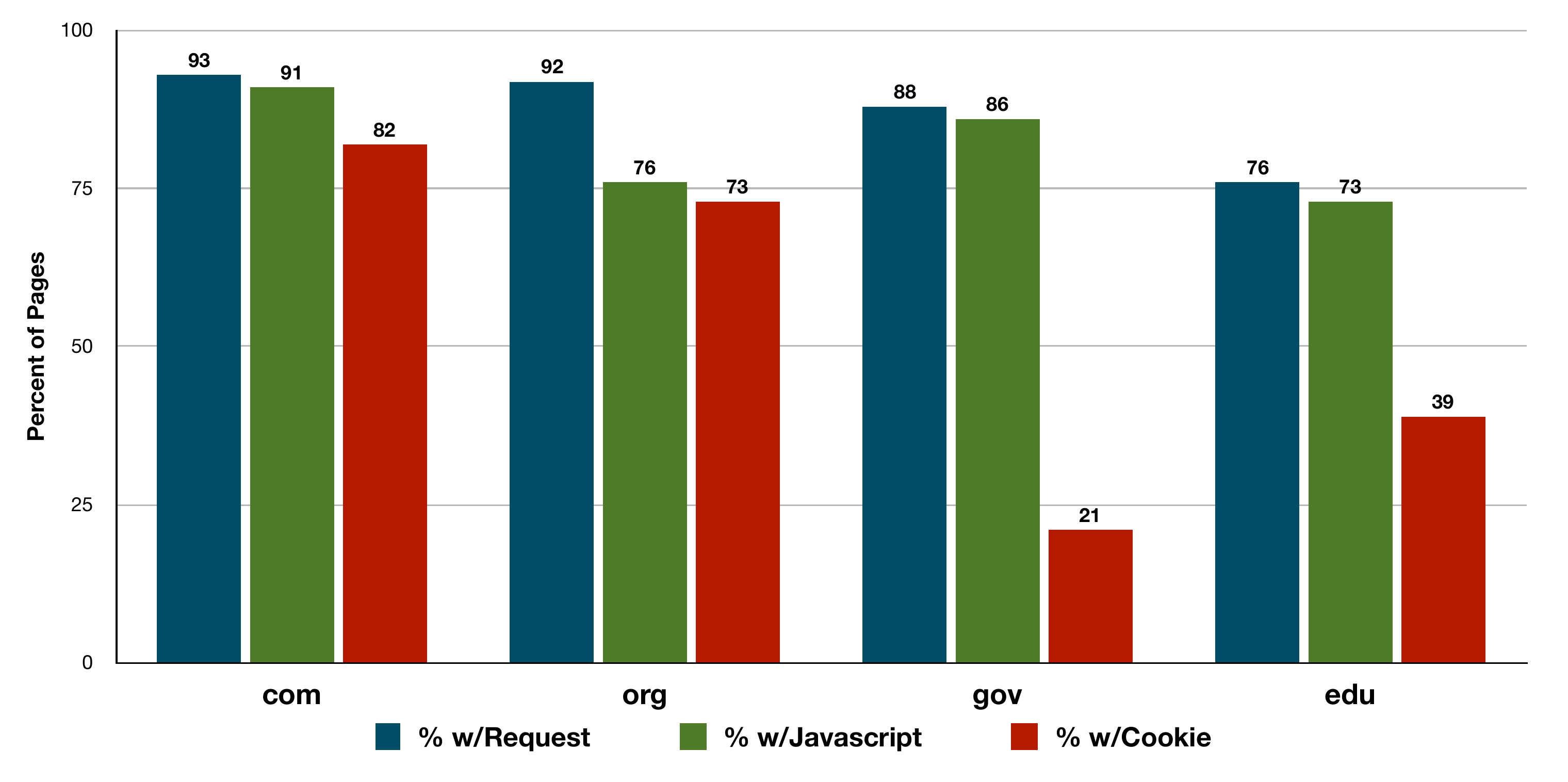}
		\caption{Prevalence of Third-Party Requests, Javascript, and Cookies by TLD}
	\end{figure*}

	\subsection{Request Analysis}
	The primary goal of \emph{webXray} is to identify third-party requests by comparing the domain of the web page being visited to the domains of requests being made.  For example, the address ``http://example.com/'' and the request ``http://images.example.com/logo.png'' both share the domain ``example.com", thus constituting a first-party request.  Alternately, a request from the same page to ``http://www.google-analytics.com/ga.js'', which has the domain ``google-analytics.com'', is recognized as a third-party request.  The same technique for HTTP requests is also applied towards evaluating the presence of third-party cookies.  The method is not flawless, as a given site may actually use many domains, or a subdomain may point to an outside party.  However, when evaluating these types of requests in aggregate, such problems constitute the statistical noise which is present in any large dataset.
	
%	A complexity that arises in this process comes from parsing country-code top-level domains (ccTLDs) such as ``co.uk'' or ``ac.jp''.  Whereas the previous example merely parsed the token prior to the Top Level Domain (in this case ``.com''), applying the same method to the page ``http://example.co.uk/index.html'' with the request ``http://tracker.co.uk'' would yield an incorrect match on the domain ``co.uk".  In order to cope with this, Mozilla provides a free ``Public Suffix List" which which includes most ccTLDs, allowing \emph{webXray} to recognize that domains such as ``example.co.uk" and ``trackers.co.uk" are distinct.  

	Finally, in order to evaluate larger trends in tracking mechanisms, third-party requests are dissected to extract arguments (e.g. ``?SITEID=123'') and file extensions such as .js (Javascript), .jpg (image), and .css (cascading style sheet).  Removing arguments also allows for a more robust analysis of which elements are the most prevalent, as argument strings often have specific site identifiers, making them appear unique when they are not.

	\subsection{Corporate Ownership}
	A specific focus of this investigation is to determine which corporate bodies are receiving information from health-related web pages.  While it is possible to programmatically detect requests to third-party domains, it is not always clear who the requested domains belong to.  By examining domain registration records, I have been able to pair seemingly-obscure domain names (e.g. ``2mdn.net", ``fbcdn.net") with their corporate owners (e.g. Google, Facebook).  This process has allowed me to follow the data trail back to the corporations which are the recipients of user data.  To date the literature has given much more attention to technical mechanisms, and much less to the underlying corporate dynamics.  This fresh analytical focus highlights the power of a handful of corporate giants.
	
	% revised/added section
	% per comment: "It would be good if the author could discuss the specific limitations of the used methodology."
	
	\subsection{Limitations}
		While this methodology is resource efficient and performs well at large scale, it comes with several potential limitations, many of which would produce an under-count of the number of third-party requests.  First, given the rapid rate by which pages are accessed, it is possible that rate limiting mechanisms on servers may be triggered (i.e. the requests generated by my IP would be identifiable as automated), and my IP address could be blacklisted, resulting in a under-count.  Second, due to the fact that I use PhantomJS without browser plugins such as Flash, Java, and Silverlight, some tracking mechanisms may not load or execute properly, resulting in an under-count.  Third, many tracking mechanisms are designed to be difficult to detect by a user, and an under-count could result from a failure to detect particularly clever tracking mechanisms.  Therefore, the findings presented below constitute a \emph{lower bound} of the amount of requests being made.
		
\section{Findings}
	In April 2014, I scanned 80,142 web pages which were collected from search results for  1,986 common diseases with the intent of detecting the extent and the ways in which the sensitive health data of users was being leaked.

	\subsection{General Trends}

%	\begin{table}[H]
%		\centering
%		\caption{Aggregate Trends}
%		\begin{tabular}{|l|l|c|c|c|c|c|} \hline
%			\textbf{TLD}	&	\textbf{N}		&	\textbf{\% w/Request}	
%				&	\textbf{w/JS}	&	\textbf{w/Cookie}	\\ \hline
%			*	&	80,142	&	91			&	
%					86		&	70			\\ \hline
%			com	&	49,174	&	93			&	
%					91		&	84			\\ \hline
%			org	&	16,072	&	93			&	
%					76		&	60			\\ \hline
%			gov	&	7,991	&	88			&	
%					86		&	19			\\ \hline
%			edu	&	3,444	&	75			&	
%					74		&	40			\\ \hline
%		\end{tabular}
%	\end{table}
	
	I have broken up my top-level findings into five general categories based on information gleaned from the TLDs used. They are: all pages, commercial pages (.com), non-profit pages (.org), government pages (.gov), and education-related pages (.edu).  This information is illustrated in Figure 2.  Of all pages examined, 91\% initiate some form of third-party HTTP request, 86\% download and execute third-party Javascript, and 71\% utilize cookies.  Unsurprisingly, commercial pages were above the global mean and had the most third-party requests (93\%), Javascript (91\%), and cookies (82\%).  Education pages had the least third-party HTTP requests (76\%) and Javascript (73\%), with a full quarter of pages free of third-party requests.  Government pages stood out for relatively low prevalence of third-party cookies, with only 21\% of pages storing user data in this way.  Figure 2 presents these findings in greater detail.

	\subsection{Mechanisms}
	
	Given that 91\% of pages make third-party HTTP requests, it is helpful to know what exactly is being requested.  Many third-party requests lack extensions, and when viewed in a browser display only blank pages which generate HTTP requests and may also manipulate browser caches.  Such requests accounted for 47\% of the top 100 requests and may point towards emerging trends in the ongoing contest between user preferences and tracking techniques.  The second most popular type of requested elements were Javascript files (33\%).  These files are able to execute abitrary code in a user's browser and may be used to perform fingerprinting techniques, manipulate caches and HTML5 storage, as well as initiate additional requests.  The third most popular type of content is the tried-and-true image file, which accounts for 8\% of the top requested elements.  Table 1 presents additional detail into the file extensions found.

	Given that tracking occurs on the so-called ``invisible web" it initially appears odd that so many mechanisms are images.  However, when investigating the images themselves, it is clear that they provide little indication as to who they belong to, and thus users are kept in the dark as to their purpose or presence.  An examination of the top 100 requested images determined that only 24\% contained information which would alert the user that they had initiated contact with a third party.  Many images were only a single pixel in size, and are often referred to as ``tracking pixels" as their only purpose is to initiate HTTP requests.  The most popular image, found on 45\% of pages, was a single tracking pixel with the name ``\_\_utm.gif" which is part of the Google Analytics service.  The second most popular image is the clearly identifiable Facebook ``Like" button which was found on 16\% of pages.  It is unclear how many users elect to ``Like" an illness, but Facebook is able to record page visits regardless of if a user clicks the ``Like" button, or if they even have a Facebook account in the first place.  Google and Facebook are not alone however, there are a number of companies tracking users online.

	\begin{table}[H]
		\centering
		\caption{Types of File Extensions}
		\begin{tabular}{|l|c|} \hline
		\textbf{Type}	&	\textbf{\%}	\\ \hline
		No Extension	&	47	\\ \hline
		Javascript		&	33	\\ \hline
		Image			&	8	\\ \hline
		Dynamic Page	&	4	\\ \hline
		Other			&	8	\\ \hline
%		Other			&	3	\\ \hline
%		Structured Data	&	2	\\ \hline
%		Font			&	1	\\ \hline
%		Static Page		&	1	\\ \hline
%		Style Sheet		&	1	\\ \hline
		\end{tabular}
	\end{table}
	
	\subsection{Corporate Ownership}
	While security and privacy research has often focused on \emph{how} user privacy is violated, insufficient attention has been given to \emph{who} is collecting user information.  The simple answer is that a variety of advertising companies have developed a massive data collection infrastructure which is designed not only to avoid detection, but also ignore, counteract, or evade user attempts at limiting collection.  Despite the wide range of entities collecting user data online, a handful of privately-held American advertising firms dominate the landscape of the invisible web.
	
	78\% of pages analyzed included elements which were owned by Google.  Such elements represent a number of hosted services and use a variety of domain names: they range from traffic analytics (google-analytics.com), advertisements (doubleclick.net), hosted Javascript (googleapis.com), to videos (youtube.com).  Regardless of type of services provided, in some way all of these HTTP requests funnel information back to Google.  This means that a single company has the ability to record the web activity of a huge number of individuals seeking sensitive health-related information without their knowledge or consent.

	While Google is the gorilla in the room, they are far from alone.  Table 2 details the top 10 companies found as part of this analysis along with the rankings of two data brokers.  In second place is comScore who are found on 38\% of pages, followed by Facebook with 31\%.  It is striking that these two companies \emph{combined} are still have less reach than Google.
	
	Additionally, companies were categorized according to their type of revenue model.  80\% of the top ten companies are advertisers.  The only exceptions to this rule are Adobe and Amazon.  Adobe offers a mix of software and services, including traffic analytics.  Amazon is in the business of both consumer-retail sales as well as web hosting with the Amazon Web Services (AWS) division.  At present it is unclear if AWS data is integrated into Amazon product recommendations or deals, but the possibility exists.

	While advertisers dominate online tracking, I was also able to detect two major data brokers: Experian (5\% of pages), and Acxiom (3\% of pages).  The main business model of data brokers is to collect information about individuals and households in order to sell it to financial institutions, employers, marketers, and other entities with such interest.  ``Credit scores" provided by Experian help determine if a given individual qualifies for a loan, and if so, at what interest rate.  Given that a 2007 study revealed that ``62.1\% of all bankruptcies...were medical"\cite{himmelstein-2009-medical}, it is possible that some data brokers not only know when a given person suffered a medical-related bankruptcy, but perhaps even when they first searched for information on the ailment which caused their financial troubles.

	\subsection{Health Information Leakage}
	
	The HTTP 1.1 protocol specification warns that ``the source of a link [URI] might be private information or might reveal an otherwise private information source" and advises that ``[c]lients SHOULD NOT include a Referer header field in a (non-secure) HTTP request if the referring page was transferred with a secure protocol".\cite{rfc2616} In simpler terms, web pages that include third-party elements, but do not use secure HTTP requests, risk leaking sensitive data via the ``Referer" field.  Of the pages analyzed, only 3.24\% used secure HTTP, the rest used non-encrypted HTTP connections and thereby potentially transmitted sensitive information to third parties.  Unsurprisingly, a significant amount of sensitive information was included in URI strings.
	
	Based on a random sample of 500 URIs taken from the population of pages analyzed (N=80,142), 70\% contained information related to a specific symptom, treatment, or disease.  An example of an URI containing specific symptom information is:
	
	\begin{verbatim}
		http://www.nhs.uk/conditions/breast-lump/[...]
	\end{verbatim}

	a URI containing no such information is:

	\begin{verbatim}
		http://www.ncbi.nlm.nih.gov/pubmed/21722252
	\end{verbatim}

	Given that the former type of URI was by far the most prevalent, it may be seen that third-parties are being sent a large volume of sensitive URI strings which may be analyzed for the presence of specific diseases, symptoms, and treatments.  This type of leakage is a clear risk for those who wish to keep this information out of the hands of third-parties who may use it for unknown ends.

\section{Discussion}
	Defining privacy harms is a perennially difficult proposition. Health information, however, presents two main privacy risks which are interrelated.  The first is \emph{personal identification}, where an individual's name is publicly associated with their medical history.  The second is \emph{blind discrimination}, where an individual's name is not necessarily revealed, but they may be treated differently based on perceived medical conditions.
	
	\begin{table*}
		\centering
		\caption{Corporate Ownership and Risk Assessment (N=80,142)}
		\begin{tabular}{|c|c|l|l|c|c|} \hline
			\textbf{Rank}	&	\textbf{\% Pages}	&	\textbf{Company}		& \textbf{Revenue}		&	\textbf{Identification}	&	\textbf{Blind Discrimination}\\ \hline
			1		&	78			&	Google		& Advertising	&	X	&	X\\ \hline
			2		&	38			&	comScore	&	Advertising	&	--	&	X\\ \hline
			3		&	31			&	Facebook	& Advertising	&	X	&	X\\ \hline
			4		&	22			&	AppNexus	& Advertising	&	--	&	X\\ \hline
			5		&	18			&	Add This	& Advertising	&	--	&	X\\ \hline
			6		&	18			&	Twitter		& Advertising	&	--	&	X\\ \hline
			7		&	16			&	Quantcast	& Advertising	&	--	&	X\\ \hline
			8		&	16			&	Amazon		& Retail \& Hosting		&	--	&	X\\ \hline
			9		&	11			&	Adobe		& Software \& Services	&	--	&	X\\ \hline
			10		&	11			&	Yahoo!		& Advertising	&	--	&	X\\ \hline
			...		&	--			&	--			&	--			&	--	&	--\\ \hline
			31		&	5			&	Experian	& Data Broker	&	X	&	--\\ \hline
			...		&	--			&	--			&	--			&	--	&	--\\ \hline
			47		&	3			&	Acxiom		& Data Broker	&	X	&	--\\ \hline
		\end{tabular}
	\end{table*}
	
	\subsection{Personal Identification}
	While most people would probably consider details of their health lives to be of little interest or value to others, such details form the basis of a lucrative industry.  In 2013, the US Senate Committee on Commerce, Science and Transportation released a highly critical review of the current state of the so-called ``data broker" industry.  Data brokers collect, package, and sell information about specific individuals and households with virtually no oversight.  This data includes demographic information (ages, names and addresses), financial records, social media activity, as well as information on those who may be suffering from ``particular ailments, including Attention Deficit Hyperactivity Disorder, anxiety, depression...among others".\cite{ussenate-2013-databrokers}  One company, Medbase200, was reported as using ``proprietary models" to generate and sell lists with classifications such as ``rape sufferers", ``domestic abuse victims", and ``HIV/AIDS patients".\cite{wsj-2013-rapelist}
	
	It should also be noted that such models are not always accurate.  For example, individuals looking for information on the condition of a loved one may be falsely tagged as having the condition themselves.  This expands the scope of risk beyond the patient to include family and friends.  In other cases, an individual may be searching for health information out of general interest and end up on a data broker's list of ``sufferers" or ``patients".  Common clerical and software errors may also tag individuals with conditions they do not have.  The high potential for such errors also highlights the need for privacy protections.

	Furthermore, poorly-protected health information may be abused by criminals.  The retailer Target has used data-mining techniques to analyze customers' purchase history in order to predict which women may be pregnant in order to offer them special discounts on infant-related products.\cite{duhigg-2012-companies}  Even if shoppers and surfers are comfortable with companies collecting this data, that is no guarantee it is safe from thieves.  In 2013, 40 million credit and debit card numbers were stolen from Target.\cite{krebs-2013-target}  While a stolen credit card may be reissued, if Target's health-related data were leaked online, it could have a devastating impact on millions of people.  Merely storing personally-identifiable information on health conditions raises the potential for loss, theft, and abuse.

	\subsection{Blind Discrimination}
	Advertisers regularly promise that their methods are wholly anonymous and therefore benign, yet identification is not always required for discriminatory behavior to occur.  In 2013, Latanya Sweeney investigated the placement of online advertisements which implied a given name was associated with a criminal record.\cite{sweeney-2013-discrimination}  She found that the presence of such ads were not the result of particular names being those of criminals, but appeared based on the racial associations of the name, with ``black names" more often resulting in an implication of criminal record.  In this way extant societal injustices may be replicated through advertising mechanisms online.  Discrimination against the ill may also be replicated through the collection and use of browsing behavior.

	Data mining techniques often rely on an eclectic approach to data analysis.  In the same way that a stew is the result of many varied ingredients being mixed in the same pot, behavioral advertising is the result of many types of browsing behavior being mixed together in order to detect trends.  As with ingredients in a stew, no single piece of data has an overly deterministic impact on the outcome, but each has \emph{some} impact.  Adding a visit to a weather site in the data stew will have an outcome on the offers an user receives, but not in a particularly nefarious way.  However, once health information is added to the mix, it becomes inevitable that it will have some impact on the outcome.   As medical expenses leave many with less to spend on luxuries, these users may be segregated into ``data silos"\cite{turow-2012-dailyyou} of undesirables who are then excluded from favorable offers and prices.  This forms a subtle, but real, form of discrimination against those perceived to be ill.

	\subsection{Risk Assessment}

	Having collected data on how much tracking is taking place, how it occurs, and who is doing it, it is necessary to explicate how this constitutes a risk to users.  As noted earlier, there are two main types of harm which I call identification, and blind discrimination.  Table 2 shows a breakdown of how data collection by twelve companies (top ten and data brokers) impacts the two types of risk.  The two data brokers most obviously entail a personal identification risk as their entire business model is devoted to selling personal information.  It is unlikely they are selling raw web tracking data directly, but it may be used as part of aggregate measures which are sold.

	Despite the fact that Google does not sell user data, they do possess enough ``anonymous" data to identify many users by name.  Google offers a number of services which collect detailed personal information such as a user's personal email (Gmail), work email (Apps for Business), and physical location (Google Maps).  For those who use Google's social media offering, Google+, a real name is forcefully encouraged.  By combining the many types of information held by Google services, it would be fairly trivial for the company to match real identities to ``anonymous" web browsing data. Likewise, Facebook requires the use of real names for users, and as noted before, collects data on 31\% of pages; therefore, Facebook's collection of browsing data may also result in personal identification.  In contrast, Twitter allows for pseudonyms as well as opting-out of tracking occurring off-site.

	The potential for blind discrimination is most pronounced among advertisers.  As noted above, online advertisers use complex data models which combine many pieces of unrelated information to draw conclusions about ``anonymous" individuals.  Any advertiser who is collecting and processing health browsing data will use it in some way unless it is filtered and disposed of.

\section{Policy Implications}

		The privacy issues raised by this research are of a technical nature and invite technical solutions. These solutions often come in the form of ``add-on" software which users may install in their web browsers.  Such browser add-ons have proven effective at blocking certain types of behavioral tracking.\cite{mayer-2012-third}\cite{roesner-2012-detecting}  However, this type of solution places a burden on users and has not been broadly effective.  As measurement research has shown, tracking has only increased over the past decade despite technical efforts to reign it in.
		
%\footnote{\textcolor{red}{\sout{However, over the past decade a game of cat-and-mouse has been established between researchers and online advertisers.  Researchers will periodically discover a privacy issue on the web, investigate its prevalence, and offer a new browser add-on in order to ``fix" the issue.}}} 
		
		Purely technical solutions are problematic as they require a relatively high level of knowledge and technical expertise on the part of the user.  The user must first understand the complex nature of information flows online in order to seek out technical remedies.  Next, the user must be proficient enough to install and configure the appropriate browser add-ons.  This may seem trivial for the well-educated, but many who use the Internet have little education or training in computing.  Despite this, these users deserve to have their health privacy protected.

		Furthermore, add-ons are often unavailable on the default browsers of smartphones and tablets - making it difficult for even the highly skilled to protect their privacy.  A final reason that browser add-ons provide insufficient remedy is the fact that advertisers devote significant resources to overcoming such barriers and will always find creative ways to bypass user intent.  Thus, on one hand we have users who are poorly equipped to defend themselves with available technical measures, and on the other, highly motivated and well-funded corporations with cutting-edge technologies.

	In order to effectively tackle the issue of tracking on health-related pages, attention toward the underlying social dynamics is needed.  These dynamics are formalized by government and corporate policies.  By addressing policy issues directly, rather than combating obscure tracking techniques, we may produce durable solutions which outlast today's technology cycle.  Unfortunately, extant polices are few in number and weak in effect.
	
	\subsection{Extant Policies and Protections}
	Health information is one of the few types of personal information which has been granted special protections.  The Health Insurance Portability and Accountability Act (HIPAA)\cite{act-1996-health} is a U.S. law which stipulates how medical information may be handled, stored, and accessed.  HIPAA is \emph{not} meant to police business practices in general, rather it is tailored to those providing health-specific services such as doctors, hospitals, and insurance claims processors.  Yet, even within this realm, HIPAA provides incomplete protections.  Contrary to popular perceptions, HIPAA permits the disclosure of patient information between health providers and insurance claims processors without patient notification or consent.  HIPAA generally does not allow patients to restrict the flow of their sensitive data; therefore, extending HIPAA in the online domain does not present an effective approach to privacy protection.

%However, HIPAA allows disclosure of individual patient records without consent for treatment, payment, and healthcare operations. Since these disclosure decisions are made by the holders of the information (who have an inherent conflict of interest), HIPAA does not effectively allow patients to restrict the flow of their sensitive data. Therefore, extending HIPAA in the online domain is not an effective approach to privacy protection.

	Nevertheless, the US Federal Trade Commission (FTC) has established a ``Health Breach Notification Rule" which requires entities holding personally identifiable health records to notify users if such records have been stolen.\cite{ftc-2010-health-breach}  However, merely providing health information (rather than storing doctor's notes or prescription records) does not place a business under the jurisdiction of HIPAA or associated rules.  Many businesses that handle health information are subject to virtually no oversight and the main source of policy regarding the use of health information online comes in the form of self-regulation by the parties which stand to benefit the most from capturing user data: online advertisers.

	However, self-regulation has proven wholly insufficient.  No lesser authority than the FTC determined that ``industry efforts to address privacy through self-regulation have been too slow, and up to now have failed to provide adequate and meaningful protection".\cite{ftc-2010-protecting}  When self-regulations are present, there are no serious sanctions for violating the rules which advertisers draw up among themselves.  Nevertheless, the Network Advertising Initiative (NAI) has produced a ``Code of Conduct" which requires opt-in consent for advertisers to use ``precise information" about health conditions such as cancer and mental-health.\cite{nai-2013-conduct}  Yet the same policy also states that ``member companies may seek to target users on the basis of such general health categories as headaches".\cite{nai-2013-conduct}  Given that the range of ailments between cancer and a headache is incredibly broad, this directive provides virtually no oversight.  Likewise, the Digital Advertising Alliance (DAA) provides rules which also appear to protect health information, but legal scholars have determined that ``an Internet user searching for information about or discussing a specific medical condition may still be tracked under the DAA's principles".\cite{hoofnagle-2012-privacy}

	\subsection{Potential Interventions}

	Although this problem is complex, it is not intractable and there are several ways health privacy risks may be mitigated.  First, there is no reason for non-profits, educational institutions, or government-operated sites to be leaking sensitive user information to commercial parties.  While ad-revenue keeps commercial sites running, non-profits gain support from donors and grants.  Fixing this situation could be as simple as an internal policy directive on a per-institution basis, or as expansive as adopting language which would deny funding to institutions which leak user data.

	As for commercial-oriented sites, it is true that they rely on ad-tracking revenue.  However, regulatory and legislative bodies have the authority to draft and implement policies which would require a mandatory limitation on how long information from health-related websites could be retained and how it could be used.  Such policy initiatives could have significant impact, and would reflect the preferences of the public. 

	Finally, talented engineers may devote a portion of the time they spend analyzing data to developing intelligent filters to keep sensitive data quarantined.  The spark of change could be the result of a single engineer's ``20\% time" project.  If the mad rush to ingest ever more data is tempered with a disciplined approach to filtering out potentially sensitive data, businesses and users may both benefit equally.

\section{Conclusion}

	Proving privacy harms is always a difficult task.  However, this study has demonstrated that data on health information seeking is being collected by an array of entities which are not subject to regulation or oversight.  Health information may be inadvertently misused by some companies, sold by others, or even stolen by criminals.  By recognizing that health information deserves to be treated with special care, we may mitigate what harm may already be occurring and proactively avoid future problems.

\bibliographystyle{abbrv}

\begin{thebibliography}{}

\end{thebibliography}


\begin{thebibliography}{10}

\bibitem{acar-2013-fpdetective}
G.~Acar, M.~Juarez, N.~Nikiforakis, C.~Diaz, S.~G{\"u}rses, F.~Piessens, and
  B.~Preneel.
\newblock Fpdetective: dusting the web for fingerprinters.
\newblock In {\em Proceedings of the 2013 ACM SIGSAC conference on Computer \&
  communications security}, pages 1129--1140. ACM, 2013.

\bibitem{ackerman-1999-privacy}
M.~S. Ackerman, L.~F. Cranor, and J.~Reagle.
\newblock Privacy in e-commerce: examining user scenarios and privacy
  preferences.
\newblock In {\em Proceedings of the 1st ACM conference on Electronic
  commerce}, pages 1--8. ACM, 1999.

\bibitem{ayenson-2011-flash}
M.~Ayenson, D.~Wambach, A.~Soltani, N.~Good, and C.~Hoofnagle.
\newblock Flash cookies and privacy ii: Now with html5 and etag respawning.
\newblock {\em Available at SSRN 1898390}, 2011.

\bibitem{Castellucia-2013-dataharvesting2}
C.~Castellucia, S.~Grumbach, L.~Olejnik, et~al.
\newblock Data harvesting 2.0: from the visible to the invisible web.
\newblock In {\em The Twelfth Workshop on the Economics of Information
  Security}, 2013.

\bibitem{duhigg-2012-companies}
C.~Duhigg.
\newblock How companies learn your secrets.
\newblock {\em The New York Times}, 16, 2012.

\bibitem{wsj-2013-rapelist}
E.~D.~E. Dwoskin.
\newblock Data broker removes rape-victims list after journal inquiry.
\newblock {\em Wall Street Journal}, 2013.

\bibitem{eckersley-2010-unique}
P.~Eckersley.
\newblock How unique is your web browser?
\newblock In {\em Privacy Enhancing Technologies}, pages 1--18. Springer, 2010.

\bibitem{rfc2616}
R.~Fielding, J.~Gettys, J.~Mogul, H.~Frystyk, L.~Masinter, P.~Leach, and
  T.~Berners-Lee.
\newblock Hypertext transfer protocol -- http/1.1, 1999.

\bibitem{pew-2013-health}
S.~Fox and M.~Duggan.
\newblock Health online 2013.
\newblock {\em Pew Internet and American Life Project}, 2013.

\bibitem{grimes-2000-ethics}
T.~Grimes-Gruczka, C.~Gratzer, and C.~Dialogue.
\newblock {\em Ethics: Survey of Consumer Attitudes about Health Web Sites}.
\newblock California HealthCare Foundation, 2000.

\bibitem{himmelstein-2009-medical}
D.~U. Himmelstein, D.~Thorne, E.~Warren, and S.~Woolhandler.
\newblock Medical bankruptcy in the united states, 2007: results of a national
  study.
\newblock {\em The American journal of medicine}, 122(8):741--746, 2009.

\bibitem{hoofnagle-2012-privacy}
C.~Hoofnagle, J.~Urban, and S.~Li.
\newblock Privacy and modern advertising: Most us internet users want'do not
  track'to stop collection of data about their online activities.
\newblock In {\em Amsterdam Privacy Conference}, 2012.

\bibitem{jackson-2006-protecting}
C.~Jackson, A.~Bortz, D.~Boneh, and J.~C. Mitchell.
\newblock Protecting browser state from web privacy attacks.
\newblock In {\em Proceedings of the 15th international conference on World
  Wide Web}, pages 737--744. ACM, 2006.

\bibitem{jang-2010-empirical}
D.~Jang, R.~Jhala, S.~Lerner, and H.~Shacham.
\newblock An empirical study of privacy-violating information flows in
  javascript web applications.
\newblock In {\em Proceedings of the 17th ACM conference on Computer and
  communications security}, pages 270--283. ACM, 2010.

\bibitem{krebs-2013-target}
B.~Krebs.
\newblock Sources: Target investigating data breach.
\newblock {\em
  http://krebsonsecurity.com/2013/12/sources-target-investigating-data-breach/},
  2013.

\bibitem{krishnamurthy-2011-privacyleakage}
B.~Krishnamurthy, K.~Naryshkin, and C.~Wills.
\newblock Privacy leakage vs. protection measures: the growing disconnect.
\newblock In {\em Web 2.0 Security and Privacy Workshop}, 2011.

\bibitem{krishnamurthy-2009-privacy}
B.~Krishnamurthy and C.~Wills.
\newblock Privacy diffusion on the web: a longitudinal perspective.
\newblock In {\em Proceedings of the 18th international conference on World
  wide web}, pages 541--550. ACM, 2009.

\bibitem{krishnamurthy-2006-generating}
B.~Krishnamurthy and C.~E. Wills.
\newblock Generating a privacy footprint on the internet.
\newblock In {\em Proceedings of the 6th ACM SIGCOMM conference on Internet
  measurement}, pages 65--70. ACM, 2006.

\bibitem{mayer-2012-third}
J.~R. Mayer and J.~C. Mitchell.
\newblock Third-party web tracking: Policy and technology.
\newblock In {\em Security and Privacy (SP), 2012 IEEE Symposium on}, pages
  413--427. IEEE, 2012.

\bibitem{nih-2002-hippocratic}
{National Institutes of Health, History of Medicine Division}.
\newblock Greek medicine.
\newblock {\em http://www.nlm.nih.gov/hmd/greek/greek\_oath.html}, 2002.

\bibitem{nai-2013-conduct}
{Network Advertising Initiative}.
\newblock Nai code of conduct.
\newblock {\em Network Advertising Initiative}, 201.

\bibitem{nikiforakis-2013-cookieless}
N.~Nikiforakis, A.~Kapravelos, W.~Joosen, C.~Kruegel, F.~Piessens, and
  G.~Vigna.
\newblock Cookieless monster: Exploring the ecosystem of web-based device
  fingerprinting.
\newblock In {\em IEEE Symposium on Security and Privacy}, 2013.

\bibitem{phantomjs-2013-website}
PhantomJS.
\newblock Phantomjs is a headless webkit scriptable [browser] with a javascript
  api. it has fast and native support for various web standards: Dom handling,
  css selector, json, canvas, and svg.
\newblock {\em http://phantomjs.org/}, 2013.

\bibitem{roesner-2012-detecting}
F.~Roesner, T.~Kohno, and D.~Wetherall.
\newblock Detecting and defending against third-party tracking on the web.
\newblock In {\em Proceedings of the 9th USENIX conference on Networked Systems
  Design and Implementation}, pages 12--12. USENIX Association, 2012.

\bibitem{ussenate-2013-databrokers}
{Staff of Chairman Rockefeller}.
\newblock A review of the data broker industry: Collection, use, and sale of
  consumer data for marketing purposes.
\newblock {\em US Senate}, 2013.

\bibitem{sweeney-2013-discrimination}
L.~Sweeney.
\newblock Discrimination in online ad delivery.
\newblock {\em Communications of the ACM}, 56(5):44--54, 2013.

\bibitem{turow-2012-dailyyou}
J.~Turow.
\newblock {\em The daily you: How the new advertising industry is defining your
  identity and your worth}.
\newblock Yale University Press, 2012.

\bibitem{turow-2003-americans}
J.~Turow and A.~P.~P. Center.
\newblock {\em Americans \& online privacy: The system is broken}.
\newblock Annenberg Public Policy Center, University of Pennsylvania, 2003.

\bibitem{turow-2009-americans}
J.~Turow, J.~King, C.~J. Hoofnagle, A.~Bleakley, and M.~Hennessy.
\newblock Americans reject tailored advertising and three activities that
  enable it.
\newblock {\em Available at SSRN 1478214}, 2009.

\bibitem{act-1996-health}
{United States}.
\newblock Health insurance portability and accountability act of 1996.
\newblock {\em Public Law}, 104:191, 1996.

\bibitem{ftc-2010-health-breach}
{US Federal Trade Commission}.
\newblock Complying with the ftc's health breach notification rule.
\newblock {\em
  http://www.business.ftc.gov/documents/bus56-complying-ftcs-health-breach-notification-rule},
  2010.

\bibitem{ftc-2010-protecting}
{US Federal Trade Commission}.
\newblock Protecting consumer privacy in an era of rapid change preliminary
  staff report.
\newblock {\em
  http://www.ftc.gov/sites/default/files/documents/reports/federal-trade-commission-bureau-consumer-protection-preliminary-ftc-staff-report-protecting-consumer/101201privacyreport.pdf},
  2010.

\bibitem{yen-2012-host}
T.-F. Yen, Y.~Xie, F.~Yu, R.~P. Yu, and M.~Abadi.
\newblock Host fingerprinting and tracking on the web: Privacy and security
  implications.
\newblock In {\em Proceedings of NDSS}, 2012.

\end{thebibliography}

\balancecolumns

\end{document}